\documentstyle[12pt]{article}

\newcommand{\beq}{\begin{equation}}
\newcommand{\eeq}{\end{equation}}
\newcommand{\beqa}{\begin{eqnarray}}
\newcommand{\eeqa}{\end{eqnarray}}
\newcommand{\beqar}{\begin{eqnarray*}}
\newcommand{\eeqar}{\end{eqnarray*}}
\newcommand{\al}{\alpha}
\newcommand{\be}{\beta}

\newcommand{\cD}{{\cal D}}

\newcommand{\ga}{\gamma}

\renewcommand{\l}{\ell}
\newcommand{\htau}{\hat{\tau}}

\newcommand{\T}{\Theta}

\newcommand{\ssc}{\scriptscriptstyle}
\newcommand{\eg}{{\it e.g.,}\ }
\newcommand{\ie}{{\it i.e.,}\ }
\newcommand{\labell}[1]{\label{#1}} 
\newcommand{\reef}[1]{(\ref{#1})}
\newcommand\prt{\partial}

\newcommand\ls{\ell_s}

\newcommand\tr{{\tilde r}}
\newcommand\tga{[\bar\gamma_{\ssc (p)}]}

\newcommand\TT{{\widehat T}}
\newcommand\bx{\beta_z}
\newcommand\by{\beta_y}
\newcommand\ta{\bar a} 
\newcommand\tb{\bar b} 

\newcommand\cO{{\cal O}}
\newcommand\cM{{\cal M}}

\def\IR{{\hbox{{\rm I}\kern-.2em\hbox{\rm R}}}}
\begin{document}

\thispagestyle{empty}
\rightline{\small hep-th/9903203 \hfill McGill/99-06}
\vspace*{2cm}

\begin{center}
{\bf \LARGE Stress Tensors and Casimir Energies}\\[.25em]
{\bf \LARGE in the AdS/CFT Correspondence}
\vspace*{1cm}

Robert C. Myers\footnote{E-mail: rcm@hep.physics.mcgill.ca}\\
\vspace*{0.2cm}
{\it Department of Physics, McGill University}\\
{\it Montr\'eal, QC, H3A 2T8, Canada}\\
\vspace{2cm}
ABSTRACT
\end{center}
We discuss various approaches to extracting the full stress-energy
tensor of the conformal field theory from the corresponding supergravity
solutions, within the framework of the Maldacena conjecture.
This provides a more refined probe of the AdS/CFT
correspondence. We apply these techniques in considering the
Casimir energy of the conformal field theory on a torus.
It seems that either generically the corresponding supergravity solutions
are singular (\ie involve regions of large string-scale curvatures),
or that they are largely insensitive to the boundary conditions of the
CFT on the torus.
\vfill
\setcounter{page}{0}
\setcounter{footnote}{0}
\newpage

\section{Introduction} \label{intro}

The Maldacena conjecture \cite{juan} has brought a renewed interest
in the holographic principle\cite{thooft}, which asserts that a
theory of gravity in $d$ dimensions can be described in terms of
a nongravitational theory in $d-1$ dimensions. The current
activity in string theory is focused on the AdS/CFT correspondence
\cite{ed},
which implements holography with a duality between a gravitational
theory in $d$-dimensional anti-de Sitter space and a
conformal field theory living in a $(d-1)$-dimensional ``boundary"
space. This context is best understood for a specific superstring example
with $d=5$ \cite{juan,ed,gub}.
In this case, the duality maintains an equivalence between
type IIb superstring theory on $AdS_5\times S^5$,
and ${\cal N} =4$ super-Yang-Mills
theory with gauge group $U(N)$ in four dimensions.
Further in many interesting cases, it is sufficient to
only consider the
low energy limit of the superstring theory, namely, supergravity. 

A precise formulation of the AdS/CFT correspondence is made
in equating the generating
function of the connected correlation functions in the CFT
with the string/gravity partition function on the AdS space\cite{ed,gub}.
In the approximation of classical (super)gravity,  
\beq
Z_{AdS}(\phi_i)=e^{-I(\phi_i)}=\left\langle e^{\int \phi_{0,i}\cO^i}
\right\rangle_{CFT}
\labell{geneuclid}
\eeq
where $I(\phi_i)$ is the classical (super)gravity action as a functional
of the supergravity fields, $\phi_{0,i}$ are the asymptotic
``boundary'' values
of the bulk fields $\phi_i$ up to a certain rescaling \cite{ed},
and $\cO^i$ are the dual CFT operators. Treating the ``boundary''
fields $\phi_{0,i}$ as source currents in the CFT, eq.~\reef{geneuclid}
is used in calculating the correlation functions of the operators
$\cO^i$. This framework also naturally allows one to evaluate the
expectation values of the CFT operators in terms of the asymptotic
(super)gravity fields \cite{alba,scan}.

Given that part of the duality is a theory of gravity in AdS space,
one of the bulk fields will always be the graviton. So it is
natural to ask what the role of the graviton (or metric perturbations)
are in the above construction. The appropriate current-source term
in the AdS/CFT generating function \reef{geneuclid} couples the
AdS graviton to the stress-energy tensor of the
CFT \cite{igorb,igorc},
\beq
\int d^{d-1}x\, h^{ab}\,T_{ab}\ .
\labell{sourceme}
\eeq
This coupling has been
used to investigate two- and three-point correlation functions
of the stress tensor\cite{corrt,corrg}. In particular, considering
correlations protected by supersymmetry
provides a nontrivial consistency test of the
duality between IIb supergravity on $AdS_5$ and four-dimensional
super-Yang-Mills theory \cite{igorc}.

Just as in asymptotically flat space, the energy
of an asymptotically AdS solution can be determined by
the asymptotic behavior of the metric \cite{assmet,hawkhoro}. 
In the context of the AdS/CFT correspondence, this result
has the additional interpretation that the asymptotic metric perturbations
determine the energy of the corresponding CFT state (or ensemble
of states). This element of the
correspondence was examined in ref.~\cite{positive}, where it lead to
the conjecture of a new positive energy theorem for general relativity.
In the field theory, the energy is given by $E=\int\langle
T_{tt}\rangle$, and so as in the general discussion above, one is 
considering states for which the expectation value of a particular
operator, \ie the stress-energy tensor, is nonvanishing. In fact,
the expectation value of all of the individual
components of the stress-energy can
be determined from the asymptotic metric, and this is the focus
of the present paper. Having the entire stress tensor provides a more
refined tool with which to investigate the AdS/CFT correspondence,
and we will apply it in order to extend the investigation of Casimir
energies initiated in ref.~\cite{positive}.

The remainder of the paper is organized as follows: In section 2,
we consider in
detail various techniques for calculating
the expectation value of stress-energy tensor in the CFT 
from the corresponding supergravity solutions. In section 3, we apply
these techniques to examine the Casimir energy of the CFT on a toroidal
geometry. Finally, we present a discussion of our results in section 4.

While this paper was in preparation, ref.~\cite{perk} appeared which
discusses calculating the CFT stress-energy using techniques similar to
those in section 2.2. 

\section{Stress-Energy Tensor} \label{one}

As discussed above, the stress-energy tensor provides an interesting
tool with which to study the AdS/CFT correspondence.
In the following, we consider three different approaches to
extracting the field theory stress tensor from the supergravity
solutions via: (i) asymptotically flat $p$-brane geometries,
(ii) the quasilocal energy defined by Brown and York \cite{brown}
and (iii) an expansion of the asymptotic metric with an 
appropriate choice of coordinates.

\subsection{Asymptotically Flat Geometries} \label{old}

The Maldacena conjecture \cite{juan} originally emerged out of
investigations of extended branes in string theory and M theory.
Anti-de Sitter space arises as part of the near-horizon geometry
of certain branes, \eg $AdS_{4,5,7}$ for M2-, D3- and M5-branes,
respectively \cite{juan,paul}. So we begin by considering the
supergravity solutions describing a near-extremal $p$-brane in
a $d$-dimensional, asymptotically flat spacetime. The usual formula giving the
mass of a point-like object in terms of the asymptotic metric can be extended
to give the mass per unit $p$-volume of such solutions\cite{masslu}. A simple
derivation of this result begins by considering an extended $p$-dimensional
source in the linearized gravity equations. If we assume that the
brane directions are also symmetry directions,  these results may
be further extended to yield the entire stress-energy tensor for the
$p$-brane world-volume:
\beq
T_{ab}={1\over16\pi G_{ d}}
\oint d\Omega_{d-p-2}r^{d-p-2}\,n^i\left[\eta_{ab}\left(\partial_ih^c{}_c
+\partial_ih^j{}_j-\partial_jh^j{}_i\right)-\partial_ih_{ab}\right]
\labell{newlu}
\eeq
where $n^i$ is a radial unit vector in the transverse subspace, while
$h_{\mu\nu}=g_{\mu\nu}-\eta_{\mu\nu}$ is the deviation of the
(Einstein frame)
metric from that for flat space. Note that $h_{\mu\nu}$ is not a
diffeomorphism invariant quantity, and in applying eq.~\reef{newlu}, it must
be calculated using asymptotically Cartesian coordinates.
Above, the labels $a,b = 0,1,\ldots,p$ run over the world-volume
directions, while $i,j = 1,\ldots,d-p-1$ denote the transverse directions.
For $a=b=0$, eq.~\reef{newlu} reduces to the standard formula for the
mass density of the $p$-brane\cite{masslu}.

As an application of this formula, let us consider a near-extremal D3-brane
for which the 10-dimensional spacetime metric is\cite{andgary}
\beqa
&&ds^2=
H^{-1/2}(-f^2\,dt^2+dx^2+dy^2+dz^2)+H^{1/2}\left({dr^2\over f^2}+r^2d\Omega_5
\right)
\nonumber\\
&&\quad{\rm with}\quad H=1+\left({\ell\over r}\right)^4
\quad{\rm and}\quad
f^2=1-\left({\mu\over r}\right)^4
\labell{metric1}
\eeqa
while we fix the constant background dilaton as $e^\phi=1$.
As mentioned above, to apply
eq.~\reef{newlu}, we need to express the metric in isotropic coordinates,
at least asymptotically. That is we need to find a new radial coordinate
such that $dr/f=F(R)\,dR$ and $r=F(R)\,R.$ The final result is that
asymptotically
\beq
r\simeq R\left(1+{\mu^4\over8R^4}\right)
\labell{newrad}
\eeq
so that we may write the metric as
\beq
ds^2\simeq
H(R)^{-1/2}(-f(R)^2\,dt^2+dx^2+dy^2+dz^2)+H(R)^{1/2}F(R)^2(dx^i)^2
\labell{metric2}
\eeq
with $R^2=\sum_{i=1}^6(x^i)^2$. Given this form of the metric,
a straightforward calculation of eq.~\reef{newlu} yields
\beq
T_{ab}={\pi^2\over16G_{10}}\left[-\eta_{ab}(4\ell^4+\mu^4)+
4\mu^4\delta^0_a\delta^0_b\right]
\labell{stress1}
\eeq

Now this stress tensor may be regarded as including two contributions: (i)
those appearing with the introduction of the extremal D3-brane and (ii)
those due to excitations of the D3-brane above extremality, and hence
which vanish as $\mu\rightarrow0$. The precise nature of these
sources can be understood by expanding the Born-Infeld action
\cite{igorb,igorc,joep}, or by studying string scattering from D-branes
\cite{ours}. In the context of the
Maldacena conjecture, we are primarily interested in the latter since
they represent the contribution to the stress-energy
tensor by excitations in the world-volume field theory.
To isolate these contributions, we subtract off the extremal
contribution but in doing so we must be careful to subtract off that
for an extremal D3-brane with the same RR five-form charge as the
solution \reef{metric1} given above. Thus the appropriate extremal
stress tensor is found by first setting $\mu=0$ in eq.~\reef{stress1} and
then replacing $\ell^4\rightarrow\ell^2(\ell^4+\mu^4)^{1/2}$. 
It is most interesting to make the subtraction in the
limit\footnote{This corresponds to the decoupling limit\cite{juan}
in which $\ls\rightarrow0$ while holding $\mu/\ls^2$ and $\ell/\ls$
fixed.} that $\mu/\ell<<1$ which yields
\beqa
\Delta T_{ab}&=&{\pi^2\mu^4\over16G_{10}}\left[4\delta^0_a\delta^0_b+\eta_{ab}
\right]
\nonumber\\
&=&{\pi^2\mu^4\over16G_{10}}\pmatrix{3&0&0&0\cr
                                   0&1&0&0\cr
                                   0&0&1&0\cr
                                   0&0&0&1\cr}
\labell{stress2}
\eeqa

This result has a form characteristic of a thermal gas
of massless particles. In particular its trace vanishes,
\ie $\Delta T^a{}_a=\eta^{ab}\,\Delta T_{ab}=0$. This is in
keeping with the interpretation that this contribution arises
from a thermal gas in the super Yang-Mills theory on the 
world-volume of the D3-brane. 

Repeating the calculations for near-extremal M5- and M2-branes yields an
analogous $\Delta T_{ab}$ which is again isotropic and traceless.
Hence, this stress-energy lends
itself to the interpretation of being due to a thermal gas of massless
particles on the world-volume of these nondilatonic branes as well.
In the case of a general D$p$-brane, however, the result is isotropic
but not traceless. Rather $\Delta T^a{}_a\propto(p-3)^2$ indicating the
distinguished position of the D3-brane amongst the Dirichlet branes.
Of course, an essential difference is that for the generic D$p$-brane
the dilaton is no longer constant. In such a situation, there is an
intrinsic ambiguity in the definition of the energy (see, \eg \cite{merry})
and so one could with an appropriate conformal transformation find
a metric for which eq.~\reef{newlu} yields a traceless stress tensor.

\subsection{Quasilocal Formulation} \label{stress}

As discussed above in a more general setting, the AdS/CFT correspondence
describes a duality between gravity
in AdS spacetimes and a ``boundary'' field theory.
For any theory including Einstein gravity coupled to matter fields,
the boundary stress-energy tensor may be defined as follows
\cite{brown}\footnote{Brown and York's quasilocal stress tensor \cite{brown}
was first considered in the context of the AdS/CFT correspondence
in ref.~\cite{perk}.}:
Consider spacetime manifold $\cM$ with time-like boundary
$\prt\cM$.\footnote{In general in a Minkowski-signature spacetime,
one would expect the boundary to include space-like components
as well, but in the present context, these components will not
play a role} Denote the spacetime metric as $g_{\mu\nu}$, and $n^\mu$
is the outward-pointing normal to $\prt\cM$ normalized with
$n^\mu n_\mu=1$. The induced metric on the boundary, $\ga_{\mu\nu}
=g_{\mu\nu}-n_\mu n_\nu$, acts as a projection tensor onto $\prt\cM$.
The extrinsic curvature on $\prt\cM$ is given by
$\T_{\mu\nu}=-\ga_\mu{}^\rho\nabla_\rho n_\nu$. Now given the 
standard Einstein action including a boundary term
\beq
I={1\over16\pi G_d}\int_\cM d^dx\sqrt{-g}\,(R-2\Lambda)-
{1\over8\pi G_d}\oint_{\prt\cM}d^{d-1}x \sqrt{-\ga}\,\T
+I_{\rm matter}
\labell{action}
\eeq
the boundary stress tensor is given by\cite{brown}
\beq
\tau^{ab}\equiv{2\over\sqrt{-\ga}}{\delta I\over\delta\ga_{ab}}
={1\over8\pi G_d}\left(\T^{ab}-\ga^{ab}\,\T^c{}_c\right)
\labell{surfstress}
\eeq
where $a,b,c$ denote directions parallel to the boundary.
For a background solving the equations of motion, this stress tensor
will satisfy\cite{brown}
\beq
\cD_a\tau^{ab}=-T^{nb}
\labell{conserve}
\eeq
where the source on the right-hand side is a projection of
the matter stress-energy, $T^{nb}=n_\mu T^{\mu\nu}\ga_\nu{}^b$,
and $\cD_a$ is the covariant derivative projected onto $\prt\cM$.
Eq.~\reef{conserve} expresses the local conservation of the
boundary stress-energy up to the flow of matter energy-momentum
across the boundary into $\cM$. Due to the geometric confinement
in asymptotically AdS spacetimes, this source term
will vanish in the following. 

In the case of interest here, the boundary will be an asymptotic
surface at some large radius $R$. A technical problem with the above
definitions
for the action and the surface stress tensor is that they will both
diverge in the limit $R\rightarrow\infty$. This problem can be cured
\cite{hawkhoro}
for the action by subtracting the same contribution \reef{action}
for a reference background geometry for which metric $g^0_{\mu\nu}$
matches $g_{\mu\nu}$ asymptotically, \ie the boundary $\prt\cM$
can be embedded in the reference background such that $\ga^0_{ab}
=\ga_{ab}$.  This background subtraction
procedure produces a finite action,
$\hat{I}=I(g)-I^0(g^0)$, and further yields a finite surface
stress tensor
\beq
\htau^{ab}\equiv{2\over\sqrt{-\ga}}{\delta \hat{I}\over\delta\ga_{ab}}
=\tau^{ab}-(\tau^0)^{ab}
\labell{newstress}
\eeq

Now consider the case that the spacetime
has a Killing vector $\xi^\mu$ which is asymptotically time-like
and surface forming, and also that $\prt\cM$ is chosen
so that the Killing vector remains an isometry of the boundary, \ie
$\cD_{(a}\xi_{b)}=0$. In this situation, one can show \cite{brown} that
\beq
E(\xi)=\oint_Bd^{d-2}x \sqrt{-\ga}\,\xi^a\htau_{ab}\xi^b
\labell{energy}
\eeq
with $B$, a hypersurface in $\prt\cM$ orthogonal to $\xi^a$,
is a conserved charge.
If the boundary contains other spacelike Killing vectors, the latter
can also be used to define other conserved charges by replacing
one of the factors of $\xi$ by a new Killing vector in eq.~\reef{energy}.
Further with the choice that
the norm $\xi^a\xi_a=-1$ on the boundary, $E(\xi)$ coincides
precisely with the standard definition of the energy\cite{brown,hawkhoro}.

As the supergravity energy should match the total energy measured
in the field theory, this definition \reef{energy}
is useful in the last step required in matching the
surface stress tensor \reef{newstress} with the expectation value of the stress
energy in the dual CFT. While the charge in eq.~\reef{energy} is finite
for asymptotically AdS spacetimes, the measure
$\sqrt{-\ga}$ is actually asymptotically divergent.
In this situation, eq.~\reef{energy} only yields a finite result because
the components of $\htau_{ab}$ vanish asymptotically. In the
AdS/CFT duality, the asymptotic
boundary geometry is related to the background geometry on
which the dual field theory lives by a conformal transformation which
also diverges asymptotically. This conformal transformation
can be accounted for by writing the stress tensor expectation value in
the field theory as follows:
\beq
\sqrt{-h}h^{ab}\langle T_{bc}\rangle=
\lim_{R\rightarrow\infty}\sqrt{-\ga}\ga^{ab}\htau_{bc}
\labell{deaf}
\eeq
where $h_{ab}$ is the background metric of the field theory.

At this point, it may be useful to examine an explicit example
in which to apply the above analysis. Hence consider the
spherically symmetric Schwarzschild-AdS metric in $d$=$p$+2
dimensions
\beqa
&&ds^2=-f(r)^2dt^2
+f(r)^{-2}dr^2+r^2d\Omega_p
\nonumber\\
&&\qquad{\rm with}\quad
f(r)^2={r^2\over\ell^2}+1-{\mu^{p+1}\over \ell^2\,r^{p-1}}\ .
\labell{bhmetric}
\eeqa
(We have chosen a slightly unusual normalization for the mass
term to facilitate comparisons with the results for the 
planar black holes below.)
The normal vector to the surface $r=R$ is
\beq 
n^\mu=f(R)\delta^\mu_r
\labell{bnorm}
\eeq
and so the nonvanishing components of the boundary metric are
\beq
\gamma_{00}=-f(R)^2
\qquad
\ga_{\ta\tb}=R^2\,\tga_{\ta\tb}
\labell{bmet}
\eeq
where $\ta$ and $\tb$ denote angular directions, and
$\tga_{\ta\tb}$ is the metric on a unit $p$-sphere. In this
simple situation, the extrinsic curvature reduces to
\beq
\Theta_{ab}=n_r\Gamma^r{}_{ab}=-{1\over2}n^r\prt_r g_{ab}
\labell{ext}
\eeq
and with a straightforward calculation, eq.~\reef{surfstress} yields
\beq
\tau_{tt}=-{p\,f^3(R)\over8\pi G_d R}
\qquad
\tau_{\ta\tb}={R\over 8\pi G_d\,f(R)}\tga_{\ta\tb}\left(p{R^2\over\ell^2}
+p-1-{p-1\over2}{\mu^{p+1}\over \ell^2\,R^{p-1}}\right)\ .
\labell{surfa}
\eeq
With this result, we see that the nonvanishing components of the boundary
stress tensor are all diverging as $R^2$ as $R\rightarrow\infty$, making
clear the necessity of the background subtraction in eq.~\reef{newstress}.
In the present case, the natural background geometry is simply
eq.~\reef{bhmetric} with $\mu=0$, which corresponds to AdS${}_{p+2}$.
In matching the boundaries, care must be taken to scale the time
coordinate in the background metric by a constant so that 
at $r=R$, we have $\gamma^0_{tt}=\gamma_{tt}$. Eq.~\reef{newstress}
then yields
\beq
\htau_{tt}={p\over16\pi G_d \ell^3}{\mu^{p+1}\over R^{p-1}}+\cdots
\qquad
\tau_{\ta\tb}={1\over 16\pi G_d\ell} {\mu^{p+1}\over R^{p-1}}\tga_{\ta\tb}
+\cdots
\labell{surfb}
\eeq
where the ellipsis denotes terms that vanish more quickly as $R\rightarrow
\infty$.
To apply eq.~\reef{deaf}, we first define the background metric for the
field theory by stripping off the divergent conformal factor from the
boundary metric \reef{bmet}: 
\beq
h_{ab}=\lim_{R\rightarrow\infty} {\ell^2\over R^2}\gamma_{ab}
=\pmatrix{-1&\ \cr\ &\ell^2\tga_{\ta\tb}\cr}\ .
\labell{bmet2}
\eeq
The field theory stress-energy then becomes
\beq
\langle T_{ab}\rangle
={\mu^{p+1}\over16\pi G_d\ell^{p+2}}\left[(p+1)\delta^0_a\delta^0_b+h_{ab}
\right]\ .
\labell{stress3}
\eeq
So again we can recognize the form characteristic of a thermal gas of
massless particles. As a check, one can easily verify that the total
energy is
\beq
E=\oint d\Omega_p \sqrt{-h}\, \langle T_{tt}\rangle=
{p\,\Omega_p\over16\pi G_d\ell^2}\mu^{p+1}
\labell{energyb}
\eeq
where $\Omega_p=2\pi^{p+1\over2}/\Gamma({p+1\over2})$ is the area of a unit
$p$-sphere. This result agrees precisely with that calculated previously in
ref.~\cite{positive}. One can also check that this result agrees with
the energy calculated from eq.~\reef{energy} with
$\xi^a\prt_a=f(R)^{-1/2}\prt_t$.

Another interesting example to consider are the planar
black holes described by the metric
\beq
ds^2 ={r^2\over \l^2}\left[-\left (1 -{\mu^{p+1}\over r^{p+1}}\right)
 dt^2 + (dx^{\ta})^2 \right] +
 \left(1 -{\mu^{p+1}\over  r^{p+1}}\right)^{-1} {\l^2 \over r^2}dr^2 
\labell{neext}
\eeq
where $\ta=1,\cdots, p$.
For certain values of $p$, these metrics arise in the near-horizon geometry of
near-extremal
$p$-branes (see, \eg \cite{juan}). With $\mu=0$, these metrics correspond to
AdS${}_{p+2}$ space in horospheric coordinates.
Following the calculations as above, one finds that in this case
the field theory stress tensor is
\beq
\langle T_{ab}\rangle
={\mu^{p+1}\over16\pi G_d\ell^{p+2}}\left[(p+1)\delta^0_a\delta^0_b+\eta_{ab}
\right]
\labell{stress4}
\eeq
where in this case $h_{ab}=\eta_{ab}$ is simply the flat Minkowski metric
in $p+1$ dimensions. For $p=3$, eq.~\reef{neext} is precisely the
throat geometry of a near-extremal D3-brane, and comparing this
result to the previous section, we find precise agreement between
eqs.~\reef{stress2} and \reef{stress4}, when we use the identity
$G_{10}=G_5\pi^3\ell^5$.

\subsection{``Nice'' Coordinates} \label{short}

In considering absorption of gravitons by D3-branes, one finds
that gravitons with polarizations parallel to the brane
couple to the world-volume stress tensor \cite{igorb,igorc}
\beq
I_{int}={1\over2}\int d^4x\, h^{ab}T_{ab}
\labell{interact}
\eeq
As discussed in the introduction, this coupling is actually
the current-source coupling for the graviton in the
AdS/CFT generating function \reef{geneuclid}, and has been
used to investigate correlation functions
of the field theory stress tensor\cite{corrt,corrg}. As observed in
ref.~\cite{corrg}, it is convenient to perform these calculations
in ``radiation gauge'' for which
\beq
h_{r\mu}=0
\labell{radgag}
\eeq
so that the graviton polarizations are automatically in the
boundary directions. As the graviton propagates in a higher
dimensional space than the field theory, one should understand the
non-covariant coupling \reef{sourceme} as being written 
with this gauge choice in mind. 

These observations extend to situations where one is interested in
the expectation value of the stress tensor, rather than correlation
functions. We wish to determine the stress tensor of
some supergravity solution with a given metric $g$
which is to be regarded as an excitation of a background solution with
metric $g^0$ -- that is we are again calculating the stress-energy
relative to some reference background.
Now a convenient choice of coordinates can be
found such that asymptotically for large radius
\beqa
g_{rr}-g^0_{rr} &\rightarrow&o(1/r^{p+3})
\nonumber\\
g_{ra}-g^0_{ra} &\rightarrow&o(1/r^{p+1})
\labell{concord}
\eeqa
where $o(1/r^q)$ indicates that these differences are falling
off more rapidly than the indicated power of $r$. In principle, one
could consider finding coordinates such that these differences
fall off even more rapidly, but the above behavior is sufficient
to determine the expectation value of the stress-energy. With
the above choice of coordinates, the leading asymptotic perturbations
of the metric are all in components parallel to the boundary
directions. To leading order, the line element will take the
form
\beq
ds^2=g^0_{\mu\nu}dx^\mu dx^\nu+{\TT_{ab}\over r^{p-1}}dx^a dx^b
+\ldots
\labell{petg}
\eeq
One can now read off the stress tensor from the components of the metric
perturbations: $\langle T_{ab}\rangle\propto\TT_{ab}$. The constant
of proportionality (which depends only on the spacetime dimension)
can be fixed by calculating the mass of the
solution and demanding that $\langle T_{tt}\rangle$ gives the correct
mass density.

Let us apply the above procedure to
the spherically symmetric Schwarzschild-AdS metric
as an example. The radial component of the metric
\reef{bhmetric} is
\beq
g_{rr}=f(r)^{-2}={1\over {r^2\over\ell^2}+1-{\mu^{p+1}\over \ell^2\,r^{p-1}}}
\labell{radialg}
\eeq
while using AdS space as the background: $g^0_{rr}=(r^2/\ell^2+1)^{-1}$.
Now by making a transformation $r=\tr+\alpha/\tr^p$ in
the asymptotic region, one can achieve the desired fall off in
eq.~\reef{concord}. To be precise, the $r^{-(p+3)}$ perturbation
in $g_{rr}$ is eliminated
with the choice $\alpha={\mu^{p+1}\over2(p+1)}$. Inserting this
coordinate transformation into the metric \reef{bhmetric}, and comparing
with the asymptotic behavior in eq.~\reef{petg}, one finds
\beq
\TT_{tt}={p\over p+1}{\mu^{p+1}\over\ell^2}
\qquad
\TT_{\ta\tb}={\mu^{p+1}\over p+1}\tga_{\ta\tb}
\labell{almostt}
\eeq
while $\TT_{t\ta}=0$.
In order to produce the correct energy (as given in the previous
section), the proportionality constant is fixed to be
\beq
\langle T_{ab}\rangle = {p+1\over16\pi G_d\ell^p}\TT_{ab}
\labell{propcon}
\eeq
which yields a precise agreement with the stress-energy in eq.~\reef{stress3}.
The same proportionality constant (and in fact precisely the same
transformation of the radial coordinate) in considering the planar
black holes \reef{neext} again yields the correct stress-energy
\reef{stress4} using this procedure.

In the above example, we saw the traceless form of the stress tensor
emerging naturally from this choice of coordinates, as well as
precise agreement
with the results of the previous section. In fact, one can show that
the agreement between
the present prescription and that of the previous section is quite
general. Consider a metric of the above form \reef{petg}. For some
surface of fixed radius $r=R$ in the asymptotic region, the normal becomes
\beq
n_\mu dx^\mu=\sqrt{g^0_{rr}(R)}dr
\labell{normr}
\eeq
where to simplify the calculations we have assumed that $g^0_{ra}=0$
--- these metric components can be eliminated with an appropriate
choice of coordinates for generic solutions.
The complicated part of the construction is to match the asymptotic
boundary geometries in general. That is one must find coordinates
such that
\beq
g^0_{{a}{b}}|_{\bar{r}=R}=(g^0+\delta g)_{ab}|_{r=R}
\labell{matched}
\eeq
where we have denoted the metric deviation from the background
as $\delta g_{ab}=\TT_{ab}/r^{p+1}+\ldots$.
We will assume that we can accomplish this matching by a simple
scaling of the coordinates, as in the examples considered above
--- this is a limiting assumption on the generality of the discussion.
In this case, the components of the boundary metric in the background become
\beq
g^0_{{a}{b}}(r){(g^0+\delta g)_{ab}(R)\over g^0_{ab}(R)}
\labell{matcheda}
\eeq
where above the values of $a$ and $b$ are fixed.
The extrinsic curvature of the boundary now simplifies as in eq.~\reef{ext}
to yield
\beq
\Theta_{ab}=-{1\over2\sqrt{g^0_{rr}}}(g^0_{ab,r}+\delta g_{ab,r})\ .
\labell{extg}
\eeq
For the background geometry one has
\beq
\Theta^0_{ab}=-{1\over2\sqrt{g^0_{rr}}}g^0_{ab,r}{(g^0+\delta g)_{ab}(R)
\over g^0_{ab}(R)}
\labell{extgb}
\eeq
where again $a$ and $b$ are fixed in the above formula.
Carrying out the remaining calculations and substituting
in $\delta g_{ab}=\TT_{ab}/r^{p+1}+\ldots$, one then finds that
eq.~\reef{deaf} precisely reproduces the above result
\beq
\langle T_{ab}\rangle = {p+1\over16\pi G_d\ell^p}\TT_{ab}
\labell{propconb}
\eeq

\section{Casimir Energies} \label{negative}

Now we apply the results of the previous section in a discussion
of the Casimir stress-energy of the CFT on a torus. The discussion will focus
on $p=3$ for which the field theory is best understood, however, for
the most part the analysis can be extended to arbitrary
dimensions\footnote{In particular, the five-dimensional supergravity solutions
considered below are easily generalized to other dimensions\cite{newsol}.}.
We begin with a brief review of the results of ref.~\cite{positive},
which considered Casimir energies with a single compact direction.
The investigation there focussed primarily on the AdS soliton, which
is the double analytic continuation of a planar black hole given in
eq.~\reef{neext}. The AdS soliton metric for $p=3$ is:
\beq
ds^2 ={r^2\over \l^2}\left[-dt^2+ dx^2 + dy^2 +
\left (1 -{\mu^{4}\over r^4}\right)dz^2  \right] +
 \left(1 -{\mu^{4}\over  r^{4}}\right)^{-1} {\l^2 \over r^2}dr^2 
\labell{soliton}
\eeq
Here the radial coordinate is restricted to $r\ge \mu$, and geometry is
smooth at $r=\mu$ provided that $z$ is identified with period $\beta
=\pi\ell^2/\mu$. One can calculate the energy of this configuration
relative to a periodically identified AdS${}_5$ spacetime \cite{positive}.
Using the relations\footnote{That is\cite{juan}:
$g_{YM}^2=2\pi g$, $\ell^4=4\pi gN\ell_s^4$
and $G_5=8\pi^3g^2\ell_s^8/\ell^5$ where $g$ and $\ell_s$ are the string
theory coupling and length scale, respectively.}
between the AdS supergravity parameters
and those in the CFT, which is ${\cal N}=4$ super-Yang-Mills theory
with gauge group $U(N)$, one finds that the corresponding energy
density is \cite{positive}
\beq
\langle T_{tt}\rangle_{\rm sugra}=-{\pi^2\over 8}{N^2\over\beta^4}
\labell{rhosug}
\eeq
Now this negative energy density can be thought of as the Casimir energy
that is generated in the CFT when the fermions are antiperiodic on the
circle parameterized by $z$. These asymptotic boundary conditions
arise for the supergravity fermions because the $S^1$ contracts to a 
point at $r=\mu$. The Casimir energy density
can also be calculated directly in the field
theory at weak coupling, with the result being
\beq
\langle T_{tt}\rangle_{\rm gauge}=-{\pi^2\over 6}{N^2\over\beta^4}
\labell{rhogag}
\eeq
Hence one finds that this result and the negative energy density of the
supergravity solutions only differ by an overall factor of 3/4.
The weak coupling field theory calculations readily yield not
just the energy density but also the entire stress-energy tensor
which is
\beq
\langle T_{ab}\rangle_{\rm gauge}={\pi^2\over 6}{N^2\over\beta^4}
\pmatrix{-1&0&0&0\cr 0&1&0&0\cr 0&0&1&0\cr 0&0&0&-3\cr}
\labell{stress5}
\eeq
Using any of the techniques in the previous section\footnote{Note
for the purposes of
section 2.1, that eq.~\reef{soliton} can be extended to an asymptotically
flat solution by taking a doubly analytically continued near-extremal
D3-brane solution \reef{metric1}.}, a stress tensor may also be calculated
for the AdS soliton with the result
\beq
\langle T_{ab}\rangle_{\rm sugra}={\pi^2\over 8}{N^2\over\beta^4}
\pmatrix{-1&0&0&0\cr
                                   0&1&0&0\cr
                                   0&0&1&0\cr
                                   0&0&0&-3\cr}
\labell{stress6}
\eeq

These two stress-energy tensors, \reef{stress5} and \reef{stress6},
have precisely the same
form except for a single overall factor of 3/4. This discrepancy
reflects the fact that the two results apply in different regimes of the
dual gauge theory. The supergravity results, \reef{rhosug} and \reef{stress6},
correspond to the field theory for large 't Hooft coupling, 
\ie $g_{YM}^2N>>1$, while the explicit field theory results, \reef{rhogag}
and \reef{stress5}, are calculated for zero coupling, \ie $g_{YM}^2N=0$.
One can expect that the full stress-energy interpolates smoothly
between eqs.~\reef{stress5} and \reef{stress6}
as the coupling ranges between these two extremes (see, for example,
refs.~\cite{smooth,smoothie}).

Motivated by the AdS/CFT correspondence, the authors in ref.~\cite{positive}
conjectured that the AdS soliton \reef{soliton}
is actually the minimum energy solution
with these asymptotic boundary conditions. As further evidence of this
conjecture, the authors showed that the solution \reef{soliton} is
perturbatively stable against quadratic fluctuations of the metric.
This perturbative stability actually extends to many finite deformations which
continuously vary the metric \reef{soliton} which the authors explored in
their investigations \cite{nopub}.

If more than one of the spatial coordinates were compact, \ie
one might consider the CFT in $R^2\times T^2$
or $R\times T^3$ rather than $R^3\times S^1$
as above, then a natural question one might ask if the Casimir
energy is further reduced by introducing antiperiodic boundary conditions
for the fermions around more than
one of the compact directions. It is straightforward
to repeat the weak gauge coupling calculation of the stress
energy for such generalized boundary conditions. Recall that the
${\cal N}=4$ super-Yang-Mills theory contains a $U(N)$ gauge field,
six scalars in the adjoint representation, and their superpartner
fermions. The stress-energy tensor for this theory may be found in
ref.~\cite{igorc}. To leading order in a gauge coupling expansion,
the Casimir stress tensor may
be calculated by point-splitting the fields in the stress tensor
with the appropriate free-field Green's function and then removing the vacuum
divergence before taking the limit of coincident fields\cite{bird}. The
calculation is simplified by choosing orthogonal coordinates to describe
the background geometry, \ie
\beq
h_{ab}dx^adx^b=-dt^2+dx^2+dy^2+dz^2\ .
\labell{simpcord}
\eeq
To introduce the identifications producing a torus
in the spatial part of this geometry,
we must now specify three basis vectors $\vec{v}_i$ and points are
then identified according to
\beq
(x,y,z)=(x,y,z)+n_1\vec{v}_1+n_2\vec{v}_2+n_3\vec{v}_3
\labell{ident}
\eeq
where the $n_i$ are any integers. To simplify the following discussion,
we will only consider the case of $R^2\times T^2$, in which case we
will drop the vector $\vec{v}_3$. A convenient choice
for the remaining two vectors is
\beqa
\vec{v}_1&=&(0,0,\bx)
\nonumber\\
\vec{v}_2&=&(0,\beta_y\cos\theta,\by\sin\theta)\equiv
(0,\beta_yc,\by s)
\labell{vects}
\eeqa
where we will assume $\cos\theta>0$.

The desired Green's functions may now be determined by the
method of images \cite{bird}. The point-splitting calculation
then yields as the nonvanishing components of the stress 
tensor:
\beqa
\langle T_{tt}\rangle=-\langle T_{xx}\rangle
&=&-{4N^2\over\pi^2}
\sum_{m,n=-\infty}^\infty\!\!\!\!\!\!'\ \ \ 
{[1-(-1)^q]\over[(n\bx)^2+(m\by)^2+2nms\bx\by]^2}
\labell{freese}\\
\langle T_{yy}\rangle&=&{4N^2\over\pi^2}
\sum_{m,n=-\infty}^\infty\!\!\!\!\!\!'\ \ \ 
{[1-(-1)^q][(mc\by)^2-3(n\bx+ms\by)^2]\over
[(n\bx)^2+(m\by)^2+2nms\bx\by]^3}
\nonumber\\
\langle T_{zz}\rangle&=&{4N^2\over\pi^2} 
\sum_{m,n=-\infty}^\infty\!\!\!\!\!\!'\ \ \ 
{[1-(-1)^q][(n\bx+ms\by)^2-3(mc\by)^2]\over
[(n\bx)^2+(m\by)^2+2nms\bx\by]^3}
\nonumber\\
\langle T_{yz}\rangle&=&-{16N^2\over\pi^2}
\sum_{m,n=-\infty}^\infty\!\!\!\!\!\!'\ \ \ 
{[1-(-1)^q](mc\by)(n\bx+ms\by)\over
[(n\bx)^2+(m\by)^2+2nms\bx\by]^3}
\nonumber
\eeqa
where the prime on the summations indicates that summation does
not include $(m,n)=(0,0)$. The choice of the exponent $q$ depends
on the fermion boundary conditions around the $\vec{v}_1$ and 
$\vec{v}_2$ cycles:
\beq
\matrix{q&=&0\cr
         &=&n\cr
         &=&m\cr
         &=&n+m\cr}
\qquad\qquad
\matrix{(\vec{v}_1,\vec{v}_2)&=&(+,+)\cr
                             &=&(+,-)\cr
                             &=&(-,+)\cr
                             &=&(-,-)\cr}
                             \labell{boundc}
\eeq
Of course, $\langle T_{ab}\rangle=0$ for the $(+,+)$ boundary
conditions for which supersymmetry remains unbroken. In the remaining
cases, one sees that as expected the result is traceless, \ie
$\eta^{ab}\langle T_{ab}\rangle=0$. Also for generic angles, one
has an off-diagonal contribution in $\langle T_{yz}\rangle$. However,
it is straightforward to show that this term vanishes for the
special case that $\by \sin\theta= k \bx$ for some integer $k$. In this
case, one can reorganize the calculation in terms of new orthogonal
basis vectors, 
$(\vec{v}_1,\vec{v}'_2)$ where $\vec{v}_1\cdot\vec{v}'_2=0$. So one may
assume that $\by |\sin\theta|<\bx$ without loss of generality.

We are particularly interested in the energy density, which may
be rewritten as
\beqa
\langle T_{tt}\rangle&=&-{\pi^2\over6}N^2\left({\delta_{1,-}\over\bx^4}
+{\delta_{2,-}\over\by^4}\right)
\nonumber\\
&&\qquad-{16N^2\over\pi^2}\sum_{n,m=1}^\infty
{[1-(-1)^q][ (n^2\bx^2+m^2\by^2)^2+4(nms\bx\by)^2]\over
[(n^2\bx^2+m^2\by^2)^2-4(nms\bx\by)^2]^2}
\labell{enerden}
\eeqa
where $\delta_{i,-}=0$ for periodic boundary conditions around
the $i$ cycle, and $\delta_{i,-}=1$ for antiperiodic boundary conditions.
With antiperiodic boundary conditions around the first cycle, we
recover the previous result \reef{rhogag} by taking the limit
$\by\rightarrow\infty.$ For finite $\by$, it is clear that the
extra contributions make the Casimir energy density even more
negative\footnote{Even without performing the final summation, it is
clear that individual terms in the sum are either zero or negative, and
that the total sum is finite.}. In particular, even when
the second cycle has periodic
boundary conditions, even though the second term above vanishes
the infinite sum is making a negative contribution to lower the
Casimir energy below that in the previous result.

One may now ask which dual supergravity solutions describe
these field theory configurations. From the discussion in
section 2.3, one can infer the asymptotic form of the solutions.
However, finding the full solutions of the nonlinear
supergravity equations is a difficult problem. Fortunately, solutions
which appear to describe the case where $\sin\theta=0$, \ie $\vec{v}_1\cdot
\vec{v}_2=0$, are already available in the literature \cite{newsol}.
The five-dimensional metric  may be written as
\beqa
ds^2&=&{\ell^2\,dr^2\over r^2\left(1-{\mu^4\over r^4}\right)}
+{r^2\over\ell^2}\left[\left(1-{\mu^4\over r^4}
\right)^{{1\over2}(1-\al_1-\al_2)}(-dt^2+dx^2)\right.
\nonumber\\
&&\qquad\qquad\left.+\left(1-{\mu^4\over r^4}
\right)^{\al_1\vphantom{1\over2}}dy^2
+\left(1-{\mu^4\over r^4}\right)^{\al_2\vphantom{1\over2}}dz^2\right]
\labell{newmetric}
\eeqa
while the dilaton remains constant.
The exponents in eq.~\reef{newmetric} lie on the ellipse given by
\beq
3(\al_1^2+\al_2^2)+2(\al_1\al_2-\al_1-\al_2)=1\ .
\labell{ellipse}
\eeq
It is straightforward to solve this quadratic constraint to eliminate
$\al_1$ with
\beq
\al_{1\pm}={1\over3}(1-\al_2\pm2(1+\al_2-2\al_2^2)^{1/2})\ .
\labell{quadsol}
\eeq
One may note here that if one allows $\mu^4$ to take negative values,
the solutions characterized by $(\mu^4,\al_1,\al_2)$ and
$(\tilde\mu^4,\tilde\al_1,\tilde\al_2)=(-\mu^4,{1\over2}-\al_1,
{1\over2}-\al_2)$ are identical up to a diffeomorphism. Since
$\tilde\al_{1\mp}={1\over2}-\al_{1\pm}$, one need only consider the
positive branch $\al_{1+}$ in eq.~\reef{quadsol} by including negative
values of $\mu^4$.

Eq.~\reef{newmetric} becomes the AdS soliton \reef{soliton}
for $(\al_1,\al_2)=(1,0)$ or $(0,1)$, and in the limit $\mu\rightarrow0$
the solution reduces to AdS space in horospheric coordinates. Apart from
these special cases, the geometry is singular at $r=\mu$. For example,
$R_{\mu\nu\al\be}R^{\mu\nu\al\be}\sim\mu^2/\ell^4/(r-\mu)^2$ as
$r$ approaches $\mu$.

To determine the corresponding field theory stress-energy, we use the
prescription of section 2.3. We consider the background to be AdS space,
and so asymptotically the radial part of the metric can be put in the
AdS form with the coordinate transformation
\beq
r= R\left(1+{\mu^4\over8R^4}\right)\ .
\labell{trant}
\eeq
Asymptotically the metric \reef{newmetric} becomes
\beqa
ds^2&\simeq&{\ell^2\over R^2}dR^2
+{R^2\over\ell^2}\left[\left(1+{(2\al_1+2\al_2-1)\mu^4\over4R^4}\right)
(-dt^2+dx^2)\right.
\nonumber\\
&&\qquad\qquad\left.
+\left(1+{(1-4\al_1)\mu^4\over4R^4}\right)dy^2
+\left(1+{(1-4\al_2)\mu^4\over4R^4}\right)dz^2\right]
\labell{asymptt}
\eeqa
From this asymptotic metric, one can read off the metric perturbation
and then applying eq.~\reef{propcon} yields
\beq
\langle T_{ab}\rangle ={\mu^4\over16\pi G_{5}\ell^5}
\pmatrix{-(2\al_1+2\al_2-1)&0&0&0\cr
           0&2\al_1+2\al_2-1&0&0\cr
           0&0&1-4\al_1&0\cr
           0&0&0&1-4\al_2\cr}
\labell{casden}
\eeq
which has a form reminiscent of the field theory result \reef{freese}, \ie
the stress tensor is traceless and generically 
$-\langle T_{tt}\rangle=\langle T_{xx}\rangle\ne\langle T_{yy}\rangle
\ne \langle T_{zz}\rangle$.
Since this stress tensor is diagonal, however, it seems that this
solution can only describe the situation with $\sin\theta=0$, \ie
the cycles on the torus are orthogonal.

Up to this point, no consideration has been made of identification of
the $y$ and $z$ coordinates in the supergravity background
\reef{newmetric}. To parallel the field theory calculation,
we should identify $y\sim y+\beta_y$ and $z\sim z+\bx$. In the special
case of the AdS soliton, demanding that the geometry be free of singularities
relates the periodicity of one of the coordinates to the parameter
$\mu$ (as described below eq.~\reef{soliton}). However, in the present
case with generic exponents, one cannot avoid a curvature singularity
at $r=\mu$. Therefore without understanding the stringy physics
that underlies this region of strong curvature,
there is no natural way to relate the periods to
the parameter $\mu$. Given the results for the near-extremal
D-branes  and the AdS soliton, one might expect that the ratios
of the various components of the stress-energy are the same in
the strong coupling supergravity regime as arise in the weak coupling
calculations. With this assumption for a given pair of periodicities
(and $\sin\theta=0$), one could calculate the relative size
of the components of the weak coupling stress tensor \reef{freese},
and then match these ratios in the strong coupling result
\reef{casden} with a choice of exponents. Given the infinite
double sums in eq.~\reef{freese}, we have no analytical results to
offer. However, one can examine the field theory stress tensor numerically,
and it is clear (for $\sin\theta=0$) that one can always choose the exponents
in eq.~\reef{casden} to match the overall form of $\langle T_{ab}\rangle$
in the two calculations. For example, with $(-,-)$ boundary conditions,
as $\beta_z/\beta_y$ varies from 1 to $\infty$, $\langle T_{ab}\rangle$
is matched by choosing $\alpha_{1+}$ in eq.~\reef{quadsol} and choosing
$\alpha_2$ between $(1+\sqrt{3})/4$ and 1. There remains the question
of the overall normalization of the strong coupling stress-energy,
but this seems to require a knowledge of physics at string scale
curvatures. 

\section{Discussion} \label{discuss}

The field theory stress-energy tensor provides an interesting tool
with which to study the AdS/CFT correspondence. In section 2, we
have provided a number of different approaches to calculating
$\langle T_{ab}\rangle$ for a given supergravity solution. Note that
each of the calculations presented there
requires a background solution, which essentially
defines the zero for the stress tensor. That is the calculations yield
the stress-energy of the given solution relative to the reference
background.

The presence of a background solution is useful in making
contact with earlier discussions of expectation values
of CFT operators\cite{alba}.
This discussion originally relied on considering solutions of
the linearized equations of motion around AdS space, however,
it actually extended to solutions of the full nonlinear
supergravity equations in considering D-instantons\cite{alba}.
One can (at least roughly) classify the solutions
of the linearized equations as
modes which are singular at the boundary of AdS,
and those which are singular at the interior.
The modes that are singular at the boundary are
the ones associated with the source currents for
the CFT in calculating correlation functions \cite{ed},
while those that are singular at the
interior are associated with expectation values \cite{alba}.
However the fact that the latter modes are singular (\ie reach large values)
in the interior of AdS means that one must go beyond the
linearized equations of motion to consider expectation values
in general.\footnote{The linearized modes are still useful in considering
the expectation values associated with test probes moving in the
supergravity spacetime\cite{alba,scan}.} 
The black hole solutions in eqs.~\reef{bhmetric} and \reef{neext}
are a good example in that they are
solutions of the full nonlinear (super)gravity equations
of motion. However, asymptotically these solutions approach
AdS space, and one can regard the deviations of the metric from
the AdS solution as solutions of the linearized gravity equations.
Closer examination shows that these linearized solutions correspond
to modes which become singular in the interior of AdS, which now
simply means that the full solutions enter a nonlinear regime.
With the choice of coordinates in section 2.3, one can further
match the expectation value of the stress-energy to the general calculations
discussed in ref.~\cite{alba}.

On the other hand, matching the asymptotic geometry of the (super)gravity
solutions to that of a reference background is a technical nuisance.
In fact there are solutions for which there is no natural choice of
a background solution, \eg Taub-NUT-AdS --- as discussed in \cite{nuts}.
From this point of view, Balasubramanian and Kraus \cite{perk} recently
presented a superior technique. While the basis of their calculations
is the quasilocal stress-energy \cite{brown} discussed in section 2.2, they
avoid the background subtraction by introducing a ``counterterm
subtraction,'' which only relies on the intrinsic boundary geometry of 
the solution of interest. This new technique provides a definition
of the total energy which then represents a remarkable departure
from previous investigations of gravitational energy, which appears to
be unique to asymptotically AdS spacetimes.
While relying on a background subtraction was sufficient
for our investigation of the Casimir energies on toroidal geometries
in section 3, it seems that the counterterm subtraction technique will
be essential in calculating the Casimir stress-energy
for more complicated geometries. For example, one can determine
the Casimir energy for the super-Yang-Mills theory
on $R\times S^3$ from a supergravity calculation on $AdS_5$ alone
\cite{perk}.

By considering the expectation value of the stress-energy tensor
in the dual CFT, we seem to have found an interesting interpretation
of the supergravity solutions \reef{newmetric} found in ref.~\cite{newsol}.
These solutions appear to be dual to the CFT on a two-torus with nonsupersymmetric
boundary conditions imposed on the fermions around the cycles of the
torus. As mentioned previously, the solutions provided by ref.~\cite{newsol}
are general enough that the discussion here and in section 3 can be extended to the
CFT on $T^3$ or $T^4$, as well as the AdS/CFT correspondence in higher
dimensions. However, these solutions appear to be limited to the
case where the cycles on the torus are orthogonal, \eg $\sin\theta=0$
in eq.~\reef{vects}. It may be interesting to extend this family
of solutions to include tori with non-orthogonal cycles, for which generically
the stress tensor acquires off-diagonal terms. From the discussion
of section 2.3 then, one sees that the new supergravity solutions will
include nontrivial off-diagonal metric components. Of course, the
coordinates of section 2.3 may not be the optimal choice for actually
determining the full nonlinear solutions.
It may be that considering the transformation
properties of the stress-energy under the action of $SL(d,Z)$ on $T^d$
(see \eg \cite{conft}) may be useful in trying to construct
the extended family of solutions. In any event, given such a
set of solutions, it would be interesting to understand the action
of the $SL(d,Z)$ symmetry on the supergravity spacetime. It could be that this
symmetry would be a useful tool in determining the overall
normalization of the strong coupling stress-energy without recourse
to a complete understanding of string theory in regions of strong
curvature.

We should remark that we have only found an interesting interpretation
for a subset of the solutions in eq.~\reef{newmetric}. Fixing $\alpha_2$
(and hence $\alpha_{1+}$) determines the ratio of the components of
the stress-energy \reef{casden}, and so for a certain set of
boundary conditions in the CFT, this then fixes the ratio $\beta_z/\beta_y$
from eq.~\reef{freese}. Fixing $\mu^4$ then sets the overall scale, and so
through some unknown stringy physics, this determines $\beta_z$. However,
any identifications in $y$ and $z$ are left implicit in the
supergravity solution \reef{newmetric}, and so for fixed $\alpha_2$ and
$\mu^4$, these solutions exist for arbitrary values of $\beta_y$ and
$\beta_z$. Therefore our interpretation in terms of antiperiodic fermion
boundary conditions only applies to a set of measure zero in the full
space of solutions implicitly given by eq.~\reef{newmetric}. It may be
that more creative boundary conditions may allow one to provide a
CFT interpretation for some (discrete number) of the other solutions, but
it seems unlikely that a reasonable interpretation can be found
for a generic set of parameters in this family of solutions.

This then raises the question of which singularities or
which singular solutions
are physically interesting in the context of the AdS/CFT correspondence.
One response would be that such singularities simply represent regions
of strong curvature where the low energy (super)gravity theory breaks
down, but that the singularity would be ``resolved'' in a physically
sensible way by the full theory of quantum gravity (\eg superstring-
or M-theory). It was argued in ref.~\cite{nosing} that this point of view
can not be correct, by considering negative mass Schwarzschild solution
of the Einstein equations.
The argument there applies equally well in present context with a negative
cosmological constant. If the time-like singularities at the center of
negative mass ``black hole''-AdS solutions were resolved within the full
quantum gravity theory, no stable ground state would exist. Hence it
is clear that certain singularities must be unphysical and so
not all singular solutions (or solutions with regions of strong
curvature) are physically relevant. 

In certain special cases, singular solutions may be distinguished
by being supersymmetric, and
so there may be greater merit in considering such singularities. We remind
the reader that such a solution implicitly played a role in section 3. The
background solution in that case was AdS space with periodic identifications
in the horospheric coordinates. This solution has a conical singularity
at the null surface $r=0$, which is a horizon in the absence of any
identifications. The fact that the background is supersymmetric would
appear to add weight to our assumption that string theory is able to resolve
this singularity. In the AdS/CFT context, charged black hole solutions
have an interesting interpretation \cite{recent}, but it turns out that
the corresponding supersymmetric solutions \cite{larry} are actually naked 
(null) singularities whose role remains to be determined.

In the case of the metric \reef{newmetric}, the singularity is
again a null singularity as in the two preceding examples, but the solution
is not supersymmetric.
We might add that if one evaluates the supergravity action (making
a background subtraction with AdS) the result is finite \cite{newsol}.
This comes about because the metric is a solution of the supergravity
equations, and so the curvature singularity at $r=0$ does not manifest
itself in the Ricci scalar. Divergences are present in the
curvature scalars which appear as the higher order $\al'$ corrections
to the supergravity action --- see, for example, \cite{smooth}.
Of course, to properly evaluate the action including such corrections,
one would have to construct a solution of the higher order equations of
motion. However, if the higher order terms play more than a
perturbative role (as they would near the singularity), 
one must (determine and) solve the full superstring equations of motion
to produce a consistent solution.

Thus the AdS/CFT correspondence seems to provide another situation
in which curvature singularities seem to play an interesting role
in string theory (see, for example, \cite{zing}). Given the discussion
up to this point, it appears remarkable that the case with the CFT
on $R^3\times S^1$ corresponds to a dual supergravity solution which is everywhere
smooth. However, we would now like to present an alternative point of
view which argues that this behavior should not be exceptional.
The motivation for these arguments will be cosmic censorship
\cite{censor}.

Let us begin by considering the $AdS_5$ spacetime in horospheric coordinates,
\ie eq.~\reef{soliton} with $\mu^4=0$, in which one of the spatial
coordinates is periodically identified, say $y=y+\beta_y$. Further
we expect the following conclusion about the supergravity to
be independent of any boundary conditions on the supergravity
fermions in this direction. Now introducing
a planar configuration of matter into this spacetime with sufficient
density, we expect that the system will collapse to form a black
hole which settles to a metric of the form in eq.~\reef{neext} with $p=3$.
 This expectation is prejudiced by our experience
with cosmic censorship and black hole uniqueness theorems \cite{hair}
in other settings. Distinguishing the metric component $g_{yy}$ from
the other spatial directions would be like adding ``hair'' to the
black hole and produce a singular configuration similar to the solutions
\reef{newmetric}. Cosmic censorship dictates that gravity 
avoid this solution dynamically by radiating away any such hair during
the collapse that leads to the formation of the black hole.
Thus the black hole solution \reef{neext} would appear to be
the physically relevant solution {\it independent of the period $\beta_y$!}
Given this conclusion, one would expect that a double analytic
continutation will yield the supergravity solution dual
to the CFT on $R^2\times T^2$. Certainly the calculations for the
thermal ensemble on $S^1$
and the Casimir stress-energy on $T^2$ have a common
euclidean framework, as discussed in ref.~\cite{positive}. One would thus
conclude that the AdS soliton \reef{soliton} is still the relevant dual
supergravity solution independent of any additional compactifications,
as well as of the boundary conditions imposed in those directions.

Hence independent of the period $\beta_y$, 
the expectation value of the stress tensor would be that of the
AdS soliton as given in eq.~\reef{stress6}.
In particular then one has $3\langle T_{tt}\rangle
=-3\langle T_{xx}\rangle=-3\langle T_{yy}\rangle=\langle T_{zz}\rangle$,
independent of $\beta_y$ or the boundary conditions imposed on the
$y$-cycle. Certainly this result does not match the form of the weak coupling
result (even with $\sin\theta=0$) given in eq.~\reef{freese}.
Of course, the form of the latter result does go over to that
of the supergravity calculation in the limit that $\beta_y\rightarrow\infty$.
There is no inconsistency here, of course. One would simply conclude
that the various components of the stress tensor evolve 
independently as the 't Hooft parameter increases from zero
in the explicit field theory calculation to large values where
the supergravity result applies. Thus the effects of the additional
identifications seem to be erased by the supergravity dynamics in the
large $N$ limit. 

While this scenario may be reasonable in a situation where the
periodicity of one direction is much smaller than that of any others,
it should appear problematic for periodicities of roughly the same size.
Analogous to the discussion in ref.~\cite{ed},
a possible resolution is that the supergravity partition function
\reef{geneuclid}
includes contributions from different AdS solitons in each of which one of the
compact directions with antiperiodic fermions shrinks to zero
on the interior. For a given
set of periodicities, the relative size of the contributions of the
different supergravity solutions would be determined by their
euclidean action. In the case where one of the periodicities
is much smaller than the others, the corresponding AdS soliton
dominates the partition function. When the periodicities are
roughly the same size, no single solution would be dominant.
The expectation $\langle T_{ab}\rangle$ would presumably then
be given by a weighted sum of that calculated for the individual
solutions.\footnote{It appears that with this approach that there
can be a ``phase'' transition in the thermal ensemble on a torus
from the high entropy black hole to the zero-entropy AdS soliton at low
temperatures --- see refs.~\cite{ed,recent,edd} for discussions of
similar phase transitions.}

Thus we have presented two very different pictures of the
supergravity description of the dual CFT on a torus. The first
involves singular solutions, but relies on the assumption that the
ratios of the various components of the stress tensor are preserved
as the 't Hooft coupling varies. The only evidence given to support
of this assumption was that it was observed to be true for the
CFT on $R^3\times S^1$. However, this is not a very
strong argument since in that case, the form of the
stress-energy is really determined by the symmetries of the background
geometry and the traceless property of the stress tensor.
In the second scenario, the supergravity
dynamics erases the complicated details which the field theory
boundary conditions produce at weak coupling. This picture
is motivated by cosmic censorship and black hole uniqueness which,
however, have not been well studied in the context of AdS space.
Another weakness in these arguments about the dynamics
is that while it is claimed that the post-collapse black hole should be
nonsingular, the pre-collapse configuration will almost certainly
contain a singularity at the interior due to the periodic identification
of $z$. Even if the second scenario is correct, it
seems that the gravitational abhorence of singularities
must only be a large $N$ effect in the CFT. It would be interesting
to understand the physics of these configurations at finite $N$
where it is claimed that the super-Yang-Mills theory should
resolve the singularities associated with gravitational
collapse \cite{resol}.

Finally we would like to comment on the implications of the present
results for the positive energy theorem of ref.~\cite{positive}.
We have found that there are singular supergravity solutions which
have a lower energy than the AdS soliton, which was conjectured to
be a minimum energy solution. This should not come as a surprise.
One can easily find singular solutions with arbitrarily negative
energies, such as negative mass ``black holes.'' The positive
energy theorem at
least in the form of conjectures 2 and 3 in ref.~\cite{positive}
is that the AdS soliton should be the minimum energy solution
within the space of smooth solutions. The important point that
was considered here was that some singular solutions may be
physically relevant for Type IIb superstring
theory. If this is the case, it would certainly affect conjecture
1 which was phrased in terms of the full ten-dimensional Type IIb
superstring theory. As we have discussed above though, string theory
may still choose to ignore these singular solutions and describe the
dual CFT in terms of asymptotically AdS geometries which are
everywhere smooth.

\vspace{1cm}
{\bf Acknowledgements}

I would like to thank the Institute for Theoretical Physics at UCSB
and the Aspen Center for Physics for hospitality at various stages
of this project. This research was supported by NSERC of Canada, Fonds
FCAR du Qu\'ebec and at the ITP by NSF Grant PHY94-07194. 
I would also like to acknowledge useful conversations with 
A. Chamblin, N. Constable, R. Emparan, G. Horowitz, and C.V. Johnson.

\end{document}